\documentclass[twocolumn,showpacs,preprintnumbers,amsmath,amssymb]{revtex4}

\usepackage{graphicx}% Include figure files
\usepackage{dcolumn}% Align table columns on decimal point
\usepackage{bm}% bold math

\begin{document}

\preprint{APS/123-QED}

\title{Majority Model on a network with communities}

%% Notice placement of commas and superscripts and use of &
%% in the author list

\author{ R. Lambiotte$^{1}$\email{Renaud.Lambiotte@ulg.ac.be}, M. Ausloos$^{1}$ and J.A. Ho{\l}yst$^{2}$}
\affiliation{
$^1$ Universit\'e de Li\`ege, Sart-Tilman, B-4000 Li\`ege, Belgium\\
$^2$ Warsaw University of Technology, Koszykowa 75, PL-00-661 Warszaw, Poland
}

\begin{abstract}
We focus on the majority model in a topology consisting of two coupled fully-connected networks, thereby mimicking the existence of communities in social networks. We show that a transition takes place at a value of the inter-connectivity parameter. Above this value, only symmetric solutions prevail, where both communities agree with each other and reach consensus. Below this value, in contrast, the communities can reach opposite opinions and an asymmetric state is attained. The importance of the interface between the sub-networks is shown.
\end{abstract}


\maketitle

It is well-known that social networks exhibit 
modular structure of weakly coupled clusters \cite{modular}. Namely, these networks are composed of many communities of nodes, where the nodes of the same community are highly connected, while there are few links between the nodes of different communities.  It is therefore of 
particular interest to understand how well-known models of opinion formation \cite{galam,sznajd,redner} behave on these realistic geometries. It is an obvious fact \cite{elections} that the  opinion or taste of users  may strongly differ from one cluster to another due to the lack of interactions between the communities. In an economic context, for instance, it is well-known that small cliques of core users can form a niche \cite{niche, niche2} and have a different behaviour than the majority.  As an example, one may think of {\em Mac users} who concentrate in niche markets like education, graphic arts and multimedia creation \cite{mac}. From a marketing point of view, the propagation of opinion between different communities is also an important problem, due to the growing popularity of viral marketing techniques in online social networks \cite{nicheMarketing}. 

In order to address this problem, we will focus in this Rapid Communication on the majority model \cite{redner} (MR, for majority rule) applied on a simplified geometry that mimicks  community structure. MR is defined as follows. The network is composed of  $N^T$ agents. Each of them has an opinion $\alpha$ or $\beta$ (equivalently spin) about some question. E.g. Will you vote for the Republicans or the Democrats \cite{adamic}? Do you believe in Darwin's theory? Which of these two products do you want to buy?... At each time step, $G=3$ contiguous nodes are selected (the main requirement is that $G$ is an odd number) and the agents in this selection all adopt the state of the local majority. This model rests on the tendency of social agents to copy the opinion of their neighbours/friends \cite{lam} and has been shown to lead to rich collective behaviours \cite{redner}.   
For the sake of clarity, let us first focus  on MR on a fully connected network, i.e. any pair of nodes is connected.   
When $N_T \rightarrow \infty$ , it is straightforward to show that the average total number of nodes having opinion $\alpha$, denoted $A_t$, evolves as  

\begin{eqnarray}
\label{simple}
A_{t+1}= A_t - a_t (1 - 3 a_t + 2 a_t^2),
\end{eqnarray}
 where $a_t=A_t/N^T$ is the average proportion of nodes having opinion $\alpha$. 
Eq.\ref{simple} comes from the fact that the probability that two nodes $\alpha$ ($\beta$) and one node $\beta$ ($\alpha$) are selected is $a^2 (1-a)$ ($a (1-a)^2$), so that the total contribution to the evolution of $A_t$ is
\begin{eqnarray}
 W=a^2 (1-a)-a (1-a)^2  = - a (1 - 3 a + 2 a^2).
 \end{eqnarray}
 It is easy to show that Eq.\ref{simple} possesses two stable stationary solutions $a=0$ or $a=1$,  that correspond to consensus states, i.e. all agents in the system have the same opinion. The {\em mixed} solution $a=1/2$, where nodes with different opinions coexist, is also stationary but is unstable. Let us also insist on the fact that MR does not involve {\em temperature}-like parameters that randomize the opinion dynamics.

\begin{figure*}
\includegraphics[width=5.2in]{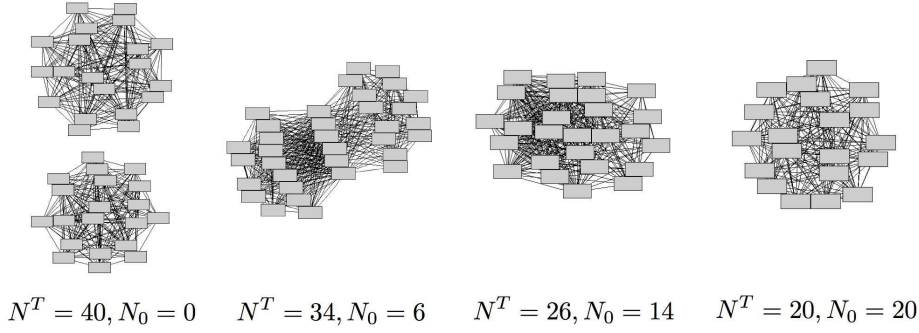}

\vspace{-0.7cm}
\caption{CFCN with $N=20$ and $\nu=0.0$, $\nu=0.3$, $\nu=0.7$ and $\nu=1.0$ (from left to right). The total number of nodes and the number of interface nodes are given for each configuration.}
\label{fig1}      
\end{figure*}

In order to generalize the fully-connected network and to account for community-like structures, we introduce coupled fully-connected networks (CFCN) defined as follows.
The system is composed of two fully connected clusters $1$ and $2$.  These two fully connected clusters are composed of $N^T_{1}$ and $N^T_{2}$ nodes respectively. The connection between the two structures is ensured by interface nodes, denoted $0$,  that belong to each of the sub-networks. By construction, the number of interface nodes verifies  $N_0 \leq min(N^T_{1},N^T_{2})$ and the case $N_0 \ll min(N^T_{1},N^T_{2})$ corresponds to a network composed of two sparsely connected communities.  In the following, we note $N_1$ and $N_2$ the number of {\em core nodes} in the clusters $1$ and $2$ respectively, where core nodes  are those that belong to only one cluster. By construction, the above quantities satisfy
\begin{eqnarray}
N_0 + N_1=N_1^T\cr
N_0 + N_2=N_2^T.
\end{eqnarray} 
For the sake of clarity, we focus on equi-populated clusters $N^T_{1}=N^T_{2}=N$.  Let us also note that the total number of nodes is $N^T=2 (1-\nu) N + \nu N=  (2-\nu) N$, where the parameter $\nu=\frac{N_0}{N}$ is a measure of the inter-connectivity between the communities. By construction, the following relations hold: $N_0= \nu N$ and $N_1=N_2= (1-\nu) N$. Some typical realizations of CFCN can be viewed in Fig.1.

In this Rapid Communication, we  will answer the following question: are there values of the inter-connectivity $\nu$ such that the co-existence of two disagreeing populations is possible?   In the limiting case $\nu \rightarrow 1$, each agent in $1$ also belongs to  $2$ and inversely, i.e. all the agents are interface agents $N_0=N^T=N$, $N_1=N_2=0$ and the network reduces to one fully connected network. Consequently, one expects that Eq.\ref{simple} takes place and that the system asymptotically reaches consensus: all the nodes either reach opinion $A$ {\bf or} opinion $B$ and co-existence is forbidden. In contrast, when $\nu=0$, the two sub-clusters are completely disconnected, $N_1=N_2=N=N^T/2$, $N_0=0$, and evolve independently from each other. Therefore, both sub-networks reach internal consensus and there is a probability $1/2$ that the opinion in 1 is the same as in 2, while these opinions differ otherwise.  The challenging problem is to find how the system behaves in the interval $\nu \in ]0,1[$.

 Before focusing on the implementation of MR on this network, let us shortly describe the above network structure. By construction, nodes in the core of 1 are connected to $N-1 \sim N$ nodes, i.e. the nodes of the core of 1 and the interface nodes, idem for the core nodes of 2. In contrast, the nodes of the interface are connected to any of the  $N^T-1 \sim N^T$ nodes in the whole network. Consequently, nodes in the core of 1 have no direct link to nodes in the core of 2, but they have an indirect connection passing through any of the interface nodes. For the sake of clarity, we will say in the following that a node is in 1 if it is in the core of 1.
 
 Let us introduce $A_{0}$,  $A_{1}$ and $A_{2}$, the average numbers of nodes $\alpha$ in the interface, in the core of 1 and in the core of 2 respectively. 
At each time step, a node is randomly chosen and a triplet of nodes centered around the chosen node is randomly selected.   By construction, the probability that a node in the core of $1$ is chosen is $p_1=\frac{N_1}{N^T}=\frac{1-\nu}{2-\nu}$. If this is the case, three possible triplets may be selected:

\begin{enumerate}
\item  The  triplet involves 3 nodes in  $1$ with probability  $p^1_{030} = (1-\nu)^2$.
\item  The  triplet involves 2 nodes in  $ 1$ and  1 node   in $0$ with probability $p^1_{120} = 2 (1-\nu) \nu$
\item  The  triplet involves 1 node in  $1$ and 2 nodes in $0$ with probability $p^1_{210} =  \nu^2$
\end{enumerate}
By convention, the  quantity $p^x_{ijk}$ is the probability that the triplet is composed of $i$ nodes in $0$, $j$ in $1$ and $k$ in $2$ ($i+j+k=3$) if the central chosen node is $x$. It respects  the normalisation $\sum_{i+j+k=3} p^x_{ijk}=1$. 

When the chosen node is in the core of $2$, with probability $p_2=\frac{N_2}{N^T}=\frac{1-\nu}{2-\nu}$, it is straightforward to get the values of $p^2_{ijk}$ by symmetry. Finally,  when the chosen node is in the interface, with probability $p_0=\frac{N_0}{N^T}=\frac{\nu}{2-\nu}$, there are six possibilities:

\begin{enumerate}
\item  The triplet involves 3 nodes in  0 with probability  $p^0_{300} = \frac{\nu^2}{(2-\nu)^2}$.
\item  The  triplet involves 2 nodes in $ 0$ and  1 node  in $1$ with probability $p^0_{210} = \frac{2 (1-\nu) \nu}{(2-\nu)^2}$
\item  The  triplet involves 2 nodes in $ 0$ and  1 node  in $2$ with probability $p^0_{201} = \frac{2 (1-\nu) \nu}{(2-\nu)^2}$
\item  The  triplet involves 1 node in $ 0$ and  2 nodes  in $1$ with probability $p^0_{120} =  \frac{(1-\nu)^2}{(2-\nu)^2}$
\item  The  triplet involves 1 node in $ 0$ and  2 nodes  in $2$ with probability $p^0_{102} =   \frac{(1-\nu)^2}{(2-\nu)^2}$
\item  The  triplet involves 1 node in each category with probability $p^0_{111} =  \frac{2 (1-\nu)^2}{(2-\nu)^2}$
\end{enumerate}

\begin{figure}
\includegraphics[width=1.5in]{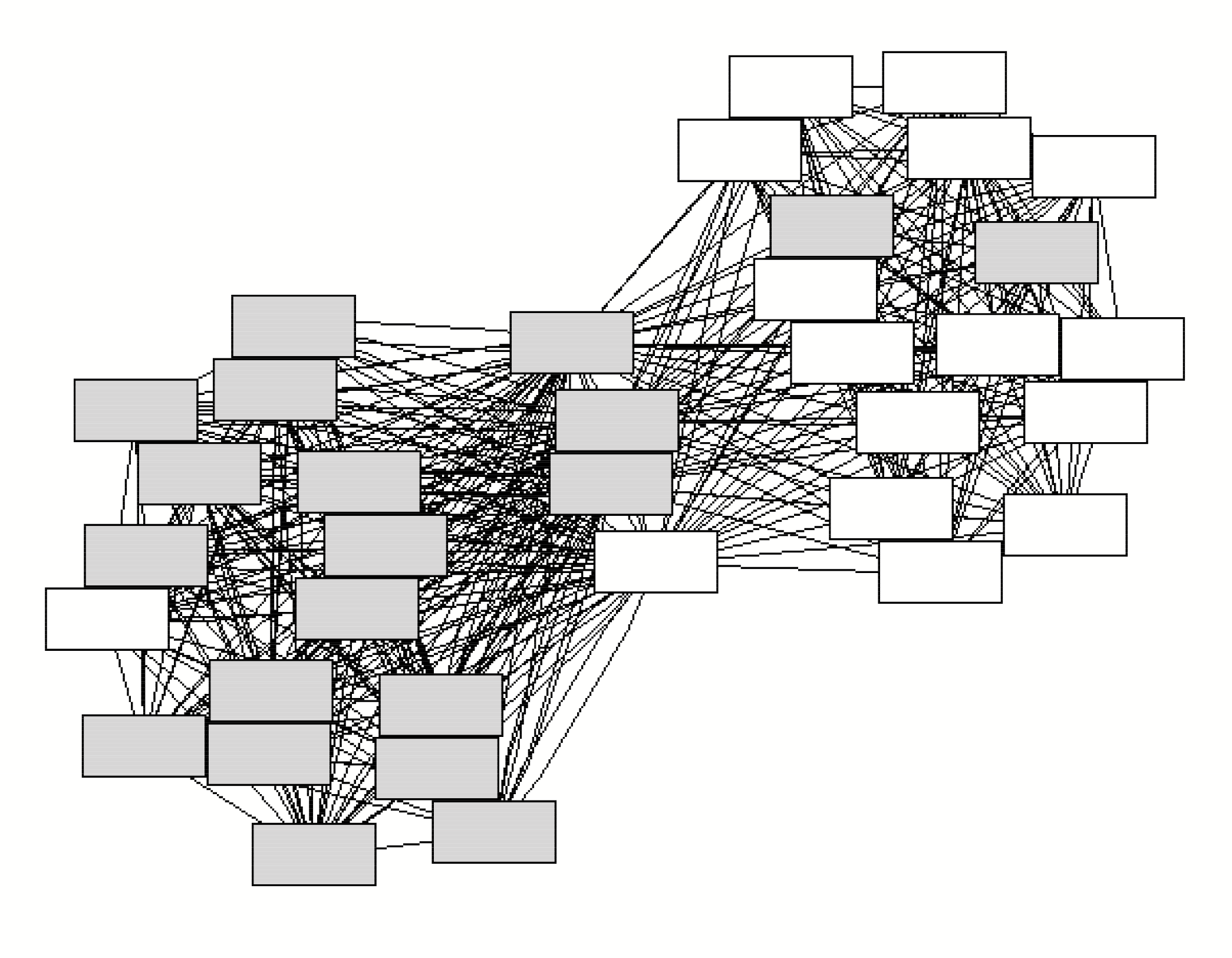}
\includegraphics[width=1.5in]{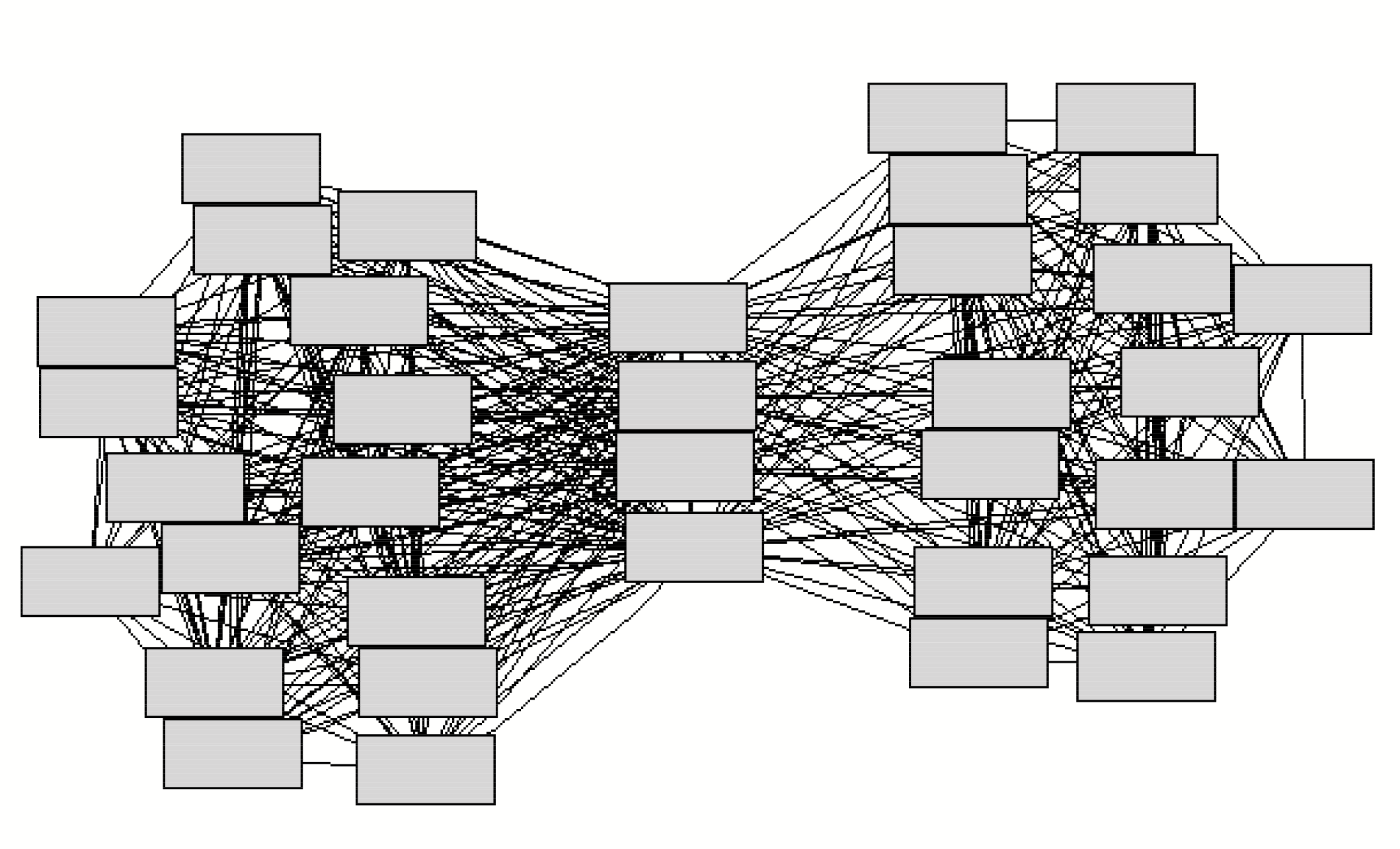}

\vspace{-0.3cm}
\caption{Typical states starting from an asymmetric initial condition $a_0=1/2$, $a_1=1$, $a_2=0$ with $N=20$ and $\nu=0.2$. Grey (white) rectangles represent nodes with opinion $\alpha$ ($\beta$). The system first reaches an asymmetric state where each cluster has a different global opinion. After some time, the system escapes the asymmetric state and reaches a  consensus state where the whole system adopts one opinion. The time for escaping the metastable state diverges when $N\rightarrow \infty$, i.e. the system remains asymmetric. The consensus state where all nodes reach opinion $\beta$ has not been plotted to avoid redundancy.}
\label{fig2}      
\end{figure}

Putting all these pieces together, the probability that a triplet $(i,j,k)$ is randomly selected  during one time step is $p_{ijk}=\sum_x p_x p^x_{ijk}$ and reads in detail:
\begin{eqnarray}
  p_{300} &=& \frac{\nu^3}{(2-\nu)^3} \cr
p_{210} &=& p_{201}  = \frac{2 (1-\nu) \nu^2}{(2-\nu)^3} + \frac{\nu^2 (1-\nu)}{(2-\nu)} \cr
&=& \nu^2 (1-\nu) \frac{6- 4 \nu + \nu^2}{(2-\nu)^3} \cr
p_{120} &=& p_{102}= \frac{(1-\nu)^2 \nu}{(2-\nu)^3} + \frac{2 \nu (1-\nu)^2}{(2-\nu)} \cr
&=& \nu (1-\nu)^2 \frac{9- 8 \nu + 2 \nu^2}{(2-\nu)^3}\cr
p_{111} &=& \frac{ 2 (1-\nu)^2 \nu}{(2-\nu)^3}\cr
p_{030} &=& p_{003} = \frac{(1-\nu)^3}{(2-\nu)}.
\end{eqnarray}

In order to derive the coupled equations generalizing Eq.\ref{simple} for quantities the $A_i$, one needs to evaluate the possible consensus processes taking place when a triplet $(i,j,k)$ is selected. Let us focus on the case $(2,1,0)$ as an example. In that case, the number of nodes $A_{0}$, $A_{1}$ and $A_{2}$ will change due to the contributions:  
\begin{eqnarray}
\label{coco}
 W_{0,(2,1,0)} &=& \frac{1}{3} [2 a_0 a_1 (1-a_0)- 2 a_0 (1-a_1) (1-a_0)]  \cr
  W_{1,(2,1,0)} &=& \frac{1}{3} [a_0^2 (1-a_1)-  a_1 (1-a_0)^2]  \cr
  W_{2,(2,1,0)} &=& 0
 \end{eqnarray}
 where the first line accounts for cases when one node 0 and one node 1 have the same opinion but disagree with a node in 0, while the second line accounts for cases when the 2 nodes in 0 have the same opinion but disagree with the node in 1. The third line simply means that the selection of a triplet $(2,1,0)$ will not change the state of a node in 2. 
The other situations $(i,j,k)$ are treated similarly. Putting all contributions together, one arrives at the set of non-linear equations: 
\begin{widetext}
 \begin{eqnarray}
 \label{complicated}
 A_{0;t+1} - A_{0;t}&=& p_{300} (a_0^2 b_0- a_0 b_0^2)+ \frac{2}{3} p_{210} (a_0 a_1 b_0- a_0 b_0 b_1)+ \frac{1}{3} p_{120} (a^2_1 b_0- a_0 b^2_1) + \frac{2}{3} p_{201} (a_0 a_2 b_0- a_0 b_0 b_2) \cr
 &+& \frac{1}{3} p_{102} (a^2_2 b_0- a_0 b^2_2)+ \frac{1}{3} p_{111} (a_1 a_2 b_0- a_0 b_1 b_2)\cr
  A_{1;t+1} - A_{1;t}&=& p_{030} (a_1^2 b_1- a_1 b_1^2)+ \frac{2}{3} p_{120} (a_0 a_1 b_1- a_1 b_0 b_1)+ \frac{1}{3} p_{210} (a^2_0 b_1- a_1 b^2_0) + \frac{1}{3} p_{111} (a_0 a_2 b_1- a_1 b_0 b_2)\cr
   A_{2;t+1} - A_{2;t}&=&  p_{003} (a_2^2 b_2- a_2 b_2^2)+ \frac{2}{3} p_{102} (a_0 a_2 b_2- a_2 b_0 b_2)+ \frac{1}{3} p_{201} (a^2_0 b_2- a_2 b^2_0) + \frac{1}{3} p_{111} (a_0 a_1 b_2- a_2 b_0 b_1)
 \end{eqnarray} 
 \end{widetext}
 where $a_i$ and $b_i$ are respectively the proportion of nodes with opinion $\alpha$ and $\beta$ in the category $i$ ($b_i=1-a_i$).
 
It is straightforward to show that $a_0=a_1=a_2=0$ or $a_0=a_1=a_2=1$ are always stationary solutions of the above coupled equations. These symmetric states correspond to  systems where the whole population has reached consensus for opinion $\alpha$ or $\beta$. However, computer simulations (Fig.2) show that an asymmetric stationary state may prevail for small enough values of $\nu$. Contrary to the symmetric state that is frozen, fluctuations continue to take place in the asymmetric state.  These fluctuations are shown to make the system escape the asymmetric  state for long enough times, i.e. it is metastable, while the absence of fluctuations in the symmetric state forbids the return to the metastable state at later times.  Computer simulations also show that the asymmetric state is characterized by averages of the form  $a_0=\frac{1}{2}$, i.e. interface nodes show no preference between $\alpha$ or $\beta$, $a_1=1/2+\epsilon$ and $a_2=1/2 - \epsilon$, where $\epsilon \in [-\frac{1}{2},\frac{1}{2}]$.
Based on these numerical results, we look for stationary solutions of Eqs.\ref{complicated} having this form. It is straightforward to show that the right hand side of the equation for $A_0$ in Eqs.\ref{complicated} is always zero in that case, while the equations for $A_1$ and $A_2$ lead to the following condition:

 \begin{figure}
\includegraphics[angle=-90,width=3.3in]{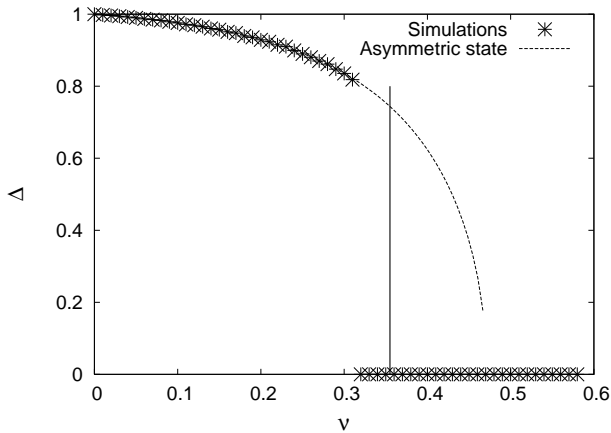}

\caption{Bifurcation diagram of $\Delta=|a_1-a_2|$ as a function of $\nu$. The simulations are performed on a network with  $N=10^6$, are started from a purely asymmetric state ($a_0=1/2$, $a_1=1$, $a_2=0$) and are stopped after $100$ steps/node. The empirical results are  in perfect agreement with the theoretical prediction, except close to the transition $\nu_{S} \sim 0.355$ (indicated by a vertical line) where the system has escaped the metastable state.  }
\label{fig1}      
\end{figure}

 \begin{eqnarray}
\epsilon \frac{(1-\nu)}{6  (2 - \nu)^2} (C+D \epsilon^2) = 0,
 \end{eqnarray} 
 where 
 $C=  6 - 17 \nu + 10 \nu^2 - 2 \nu^3$ and
$ D = -24 + 60 \nu  - 48 \nu^2 + 12 \nu^3<0$. Let us insist on the fact that Eq.7 is exact and not an expansion for small $\epsilon$.
 The trivial solution $\epsilon=0$ corresponds to an unstable state ($a_0=\frac{1}{2}$, $a_1=\frac{1}{2}$, $a_2=\frac{1}{2}$) similar to the mixed state taking place in the fully connected network. It is therefore not considered in the following. In contrast, when $C>0$, the following stationary solutions are also possible:
  \begin{eqnarray}
  \label{stationnary}
 \epsilon_{\pm} =\pm  \sqrt{-C/D}
  \end{eqnarray} 
 Solving $C>0$ numerically, one finds that the asymmetric stationary solution exists when $\nu < \nu_C$ with $\nu_C=0.471$. Values of $\nu$ for which solution Eq.\ref{stationnary} is stable and prevails in the long time limit are found by performing numerically the stability analysis \cite{nicolis} of Eqs.\ref{complicated}. To do so, one looks for solutions of the form $a_0= 1/2 + \delta_0$, $a_1= 1/2 +  \sqrt{-C/D} + \delta_1$ and $a_2= 1/2 -  \sqrt{-C/D} + \delta_2$ and keep linear terms. The linearized system evolves as $ \delta_{i;t+1} - \delta_{i;t}= \sum_{j} L_{ij} \delta_j$, where the matrix $L$ is easily found from Eqs.\ref{complicated}. We write only its first elements for the sake of readability
 
\begin{eqnarray}
L_{00} = - \frac{ (108 -328 \nu + 307 \nu^2 - 128 \nu^3 + 20 \nu^4)}{18  (2 - \nu)^3 N}   \cr
L_{01}=L_{02}= \frac{  (10 -22 \nu + 18 \nu^2 - 7 \nu^3 +  \nu^4)}{3  (2 - \nu)^3 N}\cr
...
\end{eqnarray}
By using MATHEMATICA, one finds that the critical value of $\nu$ at which one eigenvalue of $L$ becomes positive is $\nu_S \sim 0.355$. Consequently, the system exhibits a discontinuous transition at $\nu_S$: when $\nu<\nu_S$, the system may reach either the symmetric or the asymmetric solution. When $\nu > \nu_S$, only the symmetric solution is attained in the long time limit. We have performed numerical simulations of the model (Fig.3) that show an excellent agreement with the theoretical prediction Eq.\ref{stationnary}. However, there is a small discrepancy of the location of the transition: the discontinuous transition appears to take place around 0.32 in the simulations. This deviation is due to due to finite size effects, i.e. the  finite system has escaped from the metastable solution.
 Let us also stress that the above solution yields the expected value $\epsilon_{\pm} \rightarrow \pm \frac{1}{2}$  when $\nu \rightarrow 0$.

To conclude, we would like to point to the interesting features of CFCN, that allow to model topologies with well-defined communities while preserving the validity of mean field methods and allowing to identify clearly the role played by the core nodes vs. the interface nodes. Its applicability to other models relying on social networks could therefore be of interest. One may think of opinion formation (e.g. Ising models \cite{ising0,ising1,holyst2}, Voter models \cite{voter1,voter3}), language dynamics (e.g. Naming game \cite{steels1,naming})... This work could also provide a theoretical background for the use of Ising-like models in order to unravel structures in complex networks \cite{structure}. 
 
 {\bf Acknowledgements}
R.L. has been supported by European Commission Project 
CREEN FP6-2003-NEST-Path-012864.

\end{document}